\documentclass[9pt]{extarticle}
\usepackage[moderate, tracking=normal, mathdisplays=normal, charwidths=normal,leading=tight]{savetrees}
\usepackage{spconf,amsmath,graphicx}
\usepackage[utf8]{inputenc}
\usepackage{hyperref}
\usepackage{xfrac}
\usepackage[T1]{fontenc}
\usepackage{amssymb}
\usepackage{balance}
\usepackage{enumitem}
\usepackage{xcolor}
\usepackage{array}
\usepackage[font=normalsize,skip=5pt]{caption}
\usepackage[keeplastbox]{flushend}
\usepackage{mathtools}
\newcolumntype{P}[1]{>{\centering\arraybackslash}p{#1}}
\newlength{\bibitemsep}
\newlength{\bibparskip}
\let\oldthebibliography\thebibliography
\renewcommand\thebibliography[1]{%
  \oldthebibliography{#1}%
  \setlength{\parskip}{\bibitemsep}%
  \setlength{\itemsep}{\bibparskip}%
}

\makeatletter
\def\maketag@@@#1{\hbox{\m@th\normalfont\normalsize#1}}
\makeatother

\title{\llm{Massive MIMO Channel Estimation taking into account spherical waves}}
\name{\llm{Antoine Le Calvez, Luc Le Magoarou, Stéphane Paquelet} \thanks{This work has been performed in the framework of the Horizon 2020 project ONE5G (ICT-760809) receiving funds from the European Union. The authors would like to acknowledge the contributions of their colleagues in the project, although the views expressed in this contribution are those of the authors and do not necessarily represent the project.}}

\address{b\raisebox{0.2mm}{\scalebox{0.7}{\textbf{$<>$}}}com, Rennes, France}
\newcommand{\llm}[1]{{\color{black}#1}}
\begin{document}
%\ninept
%
\maketitle
%
%\nocite{*}
%\begin{multicols}{2}

\makeatletter
\let\origsection\section
\renewcommand\section{\@ifstar{\starsection}{\nostarsection}}

\newcommand\nostarsection[1]
{\sectionprelude\origsection{#1}\sectionpostlude}
\newcommand\starsection[1]
{\sectionprelude\origsection*{#1}\sectionpostlude}

\newcommand\sectionprelude{%
  \vspace{-0.3em}
}

\newcommand\sectionpostlude{%
  \vspace{-0.5em}
}
\makeatother

\makeatletter
\let\origsubsection\subsection
\renewcommand\subsection{\@ifstar{\starsubsection}{\nostarsubsection}}

\newcommand\nostarsubsection[1]
{\subsectionprelude\origsubsection{#1}\subsectionpostlude}

\newcommand\starsubsection[1]
{\subsectionprelude\origsubsection*{#1}\subsectionpostlude}

\newcommand\subsectionprelude{%
  \vspace{-0.3em}
}

\newcommand\subsectionpostlude{%
  \vspace{-0.5em}
}
\makeatother

\begin{abstract}
Together with millimiter waves (mmWaves), massive multiple-input multiple-output (MIMO) systems \llm{are key technological components} %is a technological key 
of fifth generation (5G) wireless communication systems. % to achieve its high target performance. 
In such a context, geometric considerations show that the largely adopted plane wave model (PWM) of the channel potentially loses its validity. An alternative is to consider the more accurate but more complex spherical wave model (SWM). This paper introduces an intermediate parabolic wave model (ParWM), more accurate than the PWM while less complex than the SWM. The validity domains of those three physical models are assessed in a novel way. Finally, estimation algorithms for the SWM and ParWM are proposed and compared with classical algorithms, showing a promising performance complexity trade-off.\\

\end{abstract}
\section{Introduction}
\label{sec:intro}

Massive multiple-input multiple-output (massive MIMO) is an essential technology for future fifth generation (5G) wireless communication systems
\cite{Bjornson2017, Bjornson2016, Larsson2014, Lu2014, Hampton2013, Rusek2013}. Using several antennas allows to exploit the spatial dimension to achieve high capacity, reliability, %spectral 
and energy efficiency. %These improvements are totally of interest regarding today's huge amounts of data required by the users as well as the energy consumption issues and the complex propagation environments systems are facing. 
Several Wi-Fi and 4G standards already involve classical MIMO systems typically using few antennas but the term ``massive'' refers to systems with up to hundreds of antennas with much better performance. A typical application is in cellular networks with a base station composed of many antennas and user terminals with few antennas, commonly referred to as multi-user MIMO (MU-MIMO). Massive MIMO antenna arrays are large with respect to the wavelength, so that the compactness of the system becomes a challenge. Millimiter wave (mmWave) \cite{Swindlehurst2014, Rappaport2013} operating bands mitigate this issue by reducing the wavelength.\\% if the half-wavelength antenna separation is kept as a design rule.\\
\indent The MIMO channel, assumed static and considered at a single subcarrier with $\llm{N_t}$ transmit antennas and $\llm{N_r}$ receive antennas, is usually represented in the frequency domain by the channel matrix $\mathbf{H} \in \mathbb{C}^{\llm{N_r} \times \llm{N_t}}$ containing the complex gains linking all the transmit/receive antenna couples. Knowledge of this matrix is required both at the transmitter and receiver to achieve the tremendous MIMO capacity \cite{Telatar1999}.%: it is referred to as channel state information (CSI). %The channel thus needs to be estimated, which can be performed by transmitting pilot symbols known both by the emitter and receiver. 
Estimating the entries of $\mathbf{H}$ amounts to determine $\llm{N_r} \llm{N_t}$ complex coefficients, which is not suitable in massive MIMO systems for which this number may be very high. It is thus convenient to consider a parametric channel estimation \cite{Lemagoarou2018}: injecting a priori information about the channel (combining antenna array geometry and propagation properties) allows to reduce the estimation complexity.\\
\indent Traditionally the transmitter and receiver are assumed to be separated by a large distance with respect to their antenna array size (greater than the Fraunhofer distance \cite{Selvan2017}), so that the spherical wavefronts are well approximated by planes. This simplifying hypothesis is known as the plane wave assumption, the corresponding physical model being referred to as the plane wave model (PWM). For massive MIMO systems involving up to several hundreds of antennas, i.e.\ much larger arrays, this model is not always valid and the curvature of the wavefronts cannot be neglected. In such situations, more complex but more accurate models such as the spherical wave model (SWM) might be required. 

%\begin{figure}[h]
%    \centering
%    \includegraphics[scale=0.65]{context.pdf}
%    \caption{Approximation of spherical wavefronts by planes}
%    \label{fig:context} 
%\end{figure}

\noindent \textbf{Contributions.} In this paper,
three physical parametric channel models applicable to any type of antenna array are presented. The well-known PWM and SWM are first recalled and an intermediate parabolic wave model (ParWM) is introduced, that is more accurate than the PWM while less complex than the SWM. The second contribution consists in studying the validity domains of the three models using a relative squared error metric more relevant to channel estimation than the classically used phase-shift metric.%: analytical results and simulations are provided for a line-of-sight (LoS) scenario with uniform linear and planar arrays (ULA/UPA). 
The final contribution is to propose \llm{computationally efficient channel estimation algorithms taking into account ParWM and SWM} and compare \llm{them to classical algorithms assuming the PWM.}%different estimation algorithms based on matching pursuit (MP)
%: their complexity and performance are assessed with a single-antenna receiver and multipath scenarios.
\\
\noindent \textbf{Related work.}
The PWM validity issue as well as the \llm{need for the} SWM %necessity 
to describe massive MIMO channels have already been studied in the literature \cite{Cheng2017, Tamaddondar2017, Liu2016, Liu2015, Zhou2015, Bohagen2009, Jiang2005}. So far, the studies are particularized either to linear and/or planar arrays, which yields less general and more complex analytical expressions and interpretations. In this paper the PWM and SWM analytical expressions are given adopting the generalization to any antenna array \cite{Lemagoarou2018}, and a new physical model is proposed: the ParWM. Different metrics are studied in the state of the art highlighting the PWM limits and the SWM benefits: in \cite{Cheng2017, Zhou2015} the authors investigate the correlation in single-user MIMO (SU-MIMO) and in MU-MIMO scenarios; in \cite{Cheng2017, Tamaddondar2017, Liu2016, Bohagen2009, Jiang2005} the channel capacity issue is tackled; in \cite{Selvan2017, Liu2015} the phase-shift difference induced by the SWM is used to define the near/far field boundary of large antenna arrays.  Here the models validity domains are characterized through a relative squared error metric quantifying the overall error on the channel matrix: it is discussed and compared to phase-shift considerations as presented in \cite{Selvan2017, Liu2015}. %Lastly MP based parametric channel estimation algorithms, accounting for the wave sphericity, are proposed and discussed in a massive MIMO context, which is a novel contribution to the best of the authors knowledge. 
\llm{Finally, channel estimation algorithms taking into account the curvature of the wavefronts are proposed here for the first time, to the best of the authors' knowledge.}

\section{Problem formulation}
\noindent \textbf{Notations.}
Matrices and vectors are denoted by bold upper-case and lower-case letters: $\mathbf{A}$ and $\mathbf{a}$ (except 3D "spatial" vectors that are denoted $\overrightarrow{a}$); its entry at the $i$th line and $j$th column by: $a_{ij}$.  $\mathbf{A}^T$ and $\mathbf{A}^*$ denote a matrix transpose and conjugate, respectively. The vectorization operator and the identity matrix are denoted by $\text{vec}(\cdot)$ and $\mathbf{Id}$ respectively. $\langle \cdot \rangle$, $\Vert \cdot \Vert_{\text{2}}$ and $\Vert \cdot \Vert_{\text{F}}$ denote the Hermitian inner product, the L2-norm and the Frobenius norm.

\noindent \textbf{Geometric channel model.}
\label{sec:pagestyle}
%\begin{figure}[!ht]
    %\centering
    %\includegraphics[scale=0.4]{geometrical_model.%pdf}
    %\includegraphics[scale=0.35]{geometrical_model.png}
%    \caption{Geometrical model}
%    \label{fig:geometrical_model} 
%\end{figure}
%\noindent The geometrical model used is the one introduced in \cite{Lemagoarou2018}: it is valid for any multi-antenna system in LoS. Later this geometrical model will be also used for non-LoS (NLoS) scenarios under the assumption of a single-antenna receiver ($\llm{N_r}=1$) and reflections on perfect planes. 
\llm{Let us use the channel model of \cite{Lemagoarou2018}, in which the transmit (receive) antenna array is described by the positions of its antennas denoted $\overrightarrow{a_{t,j}}, j=1,\dots,\llm{N_t}$ ($\overrightarrow{a_{r,i}}, i=1,\dots,\llm{N_r}$) with respect to its centroid $O_t$ ($O_r$). Note that the coordinate systems used at the transmitter and receiver are in general different. This description is very general: it is applicable to any type of antenna array, not only to ULA and UPA as mostly found in the literature.

 Each propagation path is described in the frequency domain by a complex gain $\rho\mathrm{e}^{\mathrm{j}\phi}$ expressing the channel between the centroids $O_t$ and $O_r$, a direction of departure (DoD) $\overrightarrow{u_t}$ (expressed in the transmitter coordinate system) and a direction of arrival $\overrightarrow{u_r}$ (expressed in the receiver coordinate system).} 
%The transmit (receive) antenna array is modeled by its centroid $O_t$ ($O_r$), a unitary wave direction of departure (arrival) $\overrightarrow{u_t}$ ($\overrightarrow{u_r}$) with azimuth $\eta_{t}$ ($\eta_{r}$), elevation $\psi_{t}$ ($\psi_{r}$), and 3D spatial vectors $\overrightarrow{a_{t,j}}$ ($\overrightarrow{a_{r,i}}$) characterizing the $\llm{N_t}$ ($\llm{N_r}$) antenna positions in a three-dimensional system of coordinates (both are not the same in general). 
\llm{Denoting} $D$ the distance between the two centroids \llm{$O_t$ and $O_r$} and $D_{ij}$ is the distance between the $j$-th transmit antenna and the $i$-th receive antenna,  
%Without loss of generality the channel coefficient for an antenna couple $ij$ can be written 
\llm{the channel for this antenna couple is classically expressed \cite{Cheng2017, Tamaddondar2017, Liu2016, Liu2015, Zhou2015, Bohagen2009, Jiang2005} as}
\begin{equation}
    \label{eq:hij_general}
    h_{ij}= \rho\mathrm{e}^{\mathrm{j}\phi}\mathrm{e}^{-\mathrm{j}\frac{2\pi}{\lambda}(D_{ij}-D)},
\end{equation}
\llm{where $\lambda$ is the wavelength and the quantity $\frac{2\pi}{\lambda}(D_{ij}-D)$ is the phase shift with respect to the reference points located at $O_t$ and $O_r$.} It has an important impact since it involves a division of the lengths difference $D_{ij}-D$ by the wavelength $\lambda$ which can be very small (1cm at 30GHz). Note that an amplitude fluctuation term also exists but is reasonably neglected since the ratio $\frac{D}{D_{ij}}$ is very close to one in practice, because the antenna arrays are in general much smaller than the propagation distance.

% entirely determined by the $D$, the antenna positions and the DoD and DoA.
%where $\rho$ and $\phi$ are the average attenuation and phase-shift, respectively. 
%The last term acts as a phase-shift fluctuation around the average one:   
%When a NLoS scenario will be considered, $\rho$ and $\phi$ will also account for the plane reflection coefficient through an additional attenuation and a phase-shift.

\noindent \textbf{Spherical Wave Model.}
\label{ssec:spherical_model}
Using the SWM consists in computing %the exact expression of 
\llm{$\Delta_{\text{SWM},ij} \triangleq D_{ij}-D$ using the channel parameters}. \llm{In a single path LoS scenario}, geometric considerations \llm{using the transmit coordinate system} lead to
\begin{equation}
    \large
    \label{eq:exact_dij}
        \resizebox{.99\hsize}{!}{$
    \begin{array}{ccl}
        \Delta_{\text{SWM},ij} & = & \left\Vert - \overrightarrow{a_{t,j}}+ D\overrightarrow{u_t} + \mathbf{R}(\delta)\overrightarrow{a_{r,i}}  \right\Vert_2-D  \\
        & & \\
         & = & D\Big(\sqrt{ 1 + \frac{2({\overrightarrow{a_{r,i}}.\overrightarrow{u_{r}} - \overrightarrow{a_{t,j}}.\overrightarrow{u_{t}}})}{D}  + \frac{\left \Vert \mathbf{R}(\delta)\overrightarrow{a_{r,i}} -  \overrightarrow{a_{t,j}} \right \Vert_2^2}{D^2}  }-1 \Big),
    \end{array}$}
\end{equation}

\normalsize 
\noindent where \llm{$- \overrightarrow{a_{t,j}}$ is the vector from the $j$-th transmit antenna to $O_t$, $D\overrightarrow{u_t}$ is the vector from $O_t$ to $O_r$ and $\mathbf{R}(\delta)\overrightarrow{a_{r,i}}$ is the vector from $O_r$ to the $i$-th receive antenna expressed in the transmit coordinate system} ($\mathbf{R}(\delta)$ is the rotation matrix mapping the receiver coordinate system to the transmit one \llm{which, given $\overrightarrow{u_t}$ and $\overrightarrow{u_r}$, depends only on a real parameter $\delta$ quantifying the rotation around the axis $O_tO_r$). The DoD and DoA being physically the same in a LoS scenario, $\mathbf{R}(\delta)\overrightarrow{u_r} = \overrightarrow{u_t}$, which allows to obtain the second line of the equation.} Note that in a single-antenna receiver case (i.e. $N_r=1$ and $\overrightarrow{a_{r,1}}=\overrightarrow{0}$) \llm{which is considered hereafter}, $\mathbf{R}(\delta)$ can be omitted. \llm{Injecting} \eqref{eq:exact_dij} in \eqref{eq:hij_general} yields the spherical channel coefficient
\begin{equation*}
    %\large
    %\label{eq:spherical_wave_model_hij}
    %\resizebox{.99\hsize}{!}{$
    h_{\text{SWM},ij}= \rho\mathrm{e}^{\mathrm{j}\phi}\mathrm{e}^{-\mathrm{j}2\pi\frac{D}{\lambda}\big(\sqrt{ 1 + \frac{2({\overrightarrow{a_{r,i}}.\overrightarrow{u_{r}} - \overrightarrow{a_{t,j}}.\overrightarrow{u_{t}}})}{D}  + \frac{\left \Vert \mathbf{R}(\delta)\overrightarrow{a_{r,i}} -  \overrightarrow{a_{t,j}} \right \Vert^2}{D^2}  }-1\big)}.
    %$}
\end{equation*}

\normalsize
\noindent Therefore the channel matrix $\mathbf{H}_{\text{SWM}}$ is a deterministic function of a set of eight parameters denoted $\boldsymbol{\theta} \triangleq \Big\{\big(\rho,\phi,\overrightarrow{u_{t}},{\overrightarrow{u_{r}}},D,{\delta}\big)\Big\}$ \llm{(two real parameters are necessary to describe each direction)}. With a large D, it is possible to perform a Taylor expansion on (\ref{eq:exact_dij}), yielding
\begin{equation}
    \label{eq:taylor_expansion}
    \resizebox{.999\hsize}{!}{$
    \begin{array}{cccl}
   \Delta_{\text{SWM},ij} & = & & \overrightarrow{a_{r,i}}.\overrightarrow{u_{r}} - \overrightarrow{a_{t,j}}.\overrightarrow{u_{t}} \\ & & \\ & & + & \frac{1}{2D}\Big[\left \Vert \mathbf{R}(\delta)\overrightarrow{a_{r,i}} -  \overrightarrow{a_{t,j}} \right \Vert^2 - (\overrightarrow{a_{r,i}}.\overrightarrow{u_{r}} - \overrightarrow{a_{t,j}}.\overrightarrow{u_{t}})^2 \Big]\\ & & \\
   %& & - & \frac{(\overrightarrow{a_{r,i}}.\overrightarrow{u_{r}} - \overrightarrow{a_{t,j}}.\overrightarrow{u_{t}})}{2D^2}\Big[\left \Vert \mathbf{R}(\delta)\overrightarrow{a_{r,i}} -  \overrightarrow{a_{t,j}} \right \Vert^2 - (\overrightarrow{a_{r,i}}.\overrightarrow{u_{r}} - \overrightarrow{a_{t,j}}.\overrightarrow{u_{t}})^2  \Big] \\ & & \\
   & & + &o\big( \frac{(R_t+R_r)^2}{2D} \big),
    \end{array}$}
\end{equation}

\normalsize
\noindent where $R_x=\underset{i}{\text{max}}\left \Vert \overrightarrow{a_{x,i}} \right \Vert$ with $x = t,r$. % The SWM \llm{thus corresponds to} an infinite order Taylor expansion. 
What if only %a \llm{finite }%limited 
 the first few orders of the expansion are considered ?

\noindent \textbf{Plane Wave Model.}
\label{ssec:plane_model}
%Actually
%Keeping only the \llm{first }%1\textsuperscript{st} 
%order, i.e.\ \llm{considering} an infinite distance $D$, yields
\llm{Approximating $\Delta_{\text{SWM},ij}$ by its first order Taylor expansion yields}
\begin{equation*}
    \label{eq:plane_wave_model_dij}
    \Delta_{\text{PWM},ij} = \overrightarrow{a_{r,i}}.\overrightarrow{u_{r}} - \overrightarrow{a_{t,j}}.\overrightarrow{u_{t}},
\end{equation*}
and leads to the well-known PWM where spherical wavefronts are approximated by planes. The PWM channel coefficient is then

\begin{equation*}
    \label{eq:plane_wave_model_hij}
    h_{\text{PWM},ij}= \rho\mathrm{e}^{\mathrm{j}\phi}\mathrm{e}^{-\mathrm{j}\frac{2\pi}{\lambda}(\overrightarrow{a_{r,i}}.\overrightarrow{u_{r}} - \overrightarrow{a_{t,j}}.\overrightarrow{u_{t}})},
\end{equation*}
and the PWM channel matrix can be expressed as
\begin{equation*}
    \label{eq:plane_wave_model_channel_matrix}
    \mathbf{H}_{\text{PWM}}= \sqrt{\llm{N_t} \llm{N_r}}{\rho}\mathrm{e}^{\mathrm{j}{\phi}} \mathbf{e}_{r}({\overrightarrow{u_r}})\mathbf{e}_{t}({\overrightarrow{u_t}})^H,
\end{equation*}
where contributions to the phase shift of the transmitter and receiver are gathered in the well-known \emph{steering vectors}
\begin{equation*}
    \label{eq:steering_vector}
    \mathbf{e}_{x}(\overrightarrow{u}) \triangleq \tfrac{1}{\sqrt{\llm{N_x}}}\begin{pmatrix}
    \mathrm{e}^{\llm{-}\mathrm{j}\frac{2\pi}{\lambda}\overrightarrow{a_{x,1}}.\overrightarrow{u}} \\
    \vdots \\
    \mathrm{e}^{\llm{-}\mathrm{j}\frac{2\pi}{\lambda}\overrightarrow{a_{x,\llm{N_x}}}.\overrightarrow{u}}
    \end{pmatrix},\;
    \text{with} \; x=t,r.
\end{equation*}
Steering vectors depend only on the direction of propagation and are insensitive to the transmission distance $D$. The PWM is thus by construction unable to take into account the curvature of the wavefronts. The PWM channel matrix $\mathbf{H}_{\text{PWM}}$ is a deterministic function of a set of six parameters denoted $\boldsymbol{\theta} \triangleq \big\{\big(\rho,\phi,\overrightarrow{u_{t}},{\overrightarrow{u_{r}}}\big)\big\}$%. The PWM is describing the channel with two parameters less than the SWM 
, which makes it less complex \llm{than the SWM} but also less accurate especially when $D$ is small, as will be shown in section~\ref{sec:validity}. % when it comes to performing estimation. 
%However one can intuitively understand that \llm{it} is less accurate: its accuracy loss will be characterized later.

\noindent \textbf{Parabolic Wave Model.}
\label{ssec:parabolic_model}
The two models presented so far are extreme: \llm{the SWM considers spherical wavefronts and the PWM approximates the spheres by planes. }%either one considers the entire sphericity of the waves or no sphericity at all. 
An intermediate solution is to \llm{approximate spheres by paraboloids} %account for some sphericity 
by considering the \llm{second} %2\textsuperscript{nd}
 order of the Taylor expansion derived in \eqref{eq:taylor_expansion}, yielding
\begin{equation*}
    \label{eq:parabolic_wave_model_dij}
    \resizebox{.99\hsize}{!}{$
    \Delta_{\text{ParWM},ij} = \overrightarrow{a_{r,i}}.\overrightarrow{u_r} - \overrightarrow{a_{t,j}}.\overrightarrow{u_t} + \frac{1}{2D}\big[\left\Vert  \mathbf{R}(\delta)\overrightarrow{a_{r,i}}- \overrightarrow{a_{t,j}}\right\Vert^2-(\overrightarrow{a_{r,i}}.\overrightarrow{u_r} - \overrightarrow{a_{t,j}}.\overrightarrow{u_t})^2\big].$}
\end{equation*}
%It is composed of
\llm{This expression comprises} the PWM term and a correction \llm{whose amplitude is} %one
 inversely proportional to the distance $D$. The ParWM channel coefficient is then
\begin{equation*}
    \label{eq:parabolic_wave_model_hij}
    \resizebox{.99\hsize}{!}{$
    h_{\text{ParWM},ij} =  \rho\mathrm{e}^{\mathrm{j}\phi}\mathrm{e}^{-\mathrm{j}\frac{2\pi}{\lambda}\Big(\overrightarrow{a_{r,i}}.\overrightarrow{u_r} - \overrightarrow{a_{t,j}}.\overrightarrow{u_t} + \frac{1}{2D}\big[\left\Vert \mathbf{R}(\delta)\overrightarrow{a_{r,i}}- \overrightarrow{a_{t,j}}\right\Vert^2-(\overrightarrow{a_{r,i}}.\overrightarrow{u_r} - \overrightarrow{a_{t,j}}.\overrightarrow{u_t})^2\big]\Big)}$}.
\end{equation*}
$\mathbf{H}_{\text{ParWM}}$ is a deterministic function of a set of eight parameters denoted $\boldsymbol{\theta} \triangleq \big\{\big(\rho,\phi,\overrightarrow{u_{t}},{\overrightarrow{u_{r}}},D,\delta\big)\big\}$. Intuitively it is obvious that the ParWM is more accurate than the PWM but less than the SWM: it will be characterized quantitatively in section~\ref{sec:validity}. However having more parameters to estimate makes it more complex than the PWM highlighting an accuracy/complexity trade-off. It has the same number of parameters as the SWM but is more tractable: its simpler expressions ease the interpretations and getting rid of the square root might have an interest for \llm{hardware implementation of} estimation algorithms.

\noindent \textbf{Single-antenna receiver \llm{and multipath channel}.}
As mentioned previously, the three physical models are valid for any $\llm{N_t}$, $\llm{N_r}$ in \llm{a single path LoS scenario}. \llm{However, further assuming a single antenna receiver ($N_r=1$ which implies $\overrightarrow{a_{r,1}} = \overrightarrow{0}$) allows to simplify the derivations. In that particular case, the above expressions are also valid for paths that originate from reflections on perfect planes \cite{Zhou2015}.} %In addition, assuming a single-antenna receiver ($\llm{N_r} = 1$) and reflections on perfect planes \llm{(as is done in \cite{Zhou2015})}, it is valid for any $\llm{N_t}$ in NLoS. 
The single-antenna receiver case is \llm{of interest} %interesting 
since it \llm{corresponds }%refers
 to a cellular network scenario with a multi-antenna base station \llm{and }% offering a connectivity to 
 multiple single-antenna user terminals \cite{Cheng2017, Tamaddondar2017, Zhou2015}: from now on the paper assumes a single-antenna receiver. It allows to derive a general expression of the channel valid for the three models in \llm{a multipath scenario} ($p$ paths) as a linear combination of \emph{characteristic vectors}:
\begin{equation}
    \label{eq:general_channel_matrix}
%    \resizebox{.99\hsize}{!}{$
    \llm{\mathbf{h}_{\mathcal{M}}}= \sqrt{\llm{N_t} }\displaystyle\sum\nolimits_{k=1}^{p}
    {\rho_k}\mathrm{e}^{\mathrm{j}{\phi_k}} \mathbf{e}_{{\mathcal{M}}}(\overrightarrow{u_{t,k}}, D_k) 
%    \left \{  
%    \begin{array}{ll} 
%    \mathcal{M}=\text{PWM, ParWM, SWM}\\
%    \boldsymbol{\tilde{\theta}}_{k,\text{PWM}} \triangleq \Big\{\big(\overrightarrow{u_{t,k}}\big)\Big\}\\
%    \boldsymbol{\tilde{\theta}}_{k,\text{ParWM}} \triangleq \Big\{\big(\overrightarrow{u_{t,k}}, D_k\big)\Big\}\\
%    \boldsymbol{\tilde{\theta}}_{k,\text{SWM}} \triangleq \Big\{\big(\overrightarrow{u_{t,k}}, D_k\big)\Big\}
%    \end{array}
%    \right. 
%$}
\end{equation}

\normalsize 
\noindent where $\overrightarrow{u_{t,k}}$ and $D_k$ are the DoD and distance of the $k$-th path, $\mathcal{M}$ denotes the considered model (PWM, ParWM or SWM), and the characteristic vector $\mathbf{e}_{{\mathcal{M}}}(\overrightarrow{u_{t,k}}, D_k)$ takes the general form %$\mathbf{e}_{\text{PWM}}(\overrightarrow{u_{t,k}}, D_k) = \mathbf{e}_{t}(\overrightarrow{u_{t,k}})^*$, %the ParWM and SWM steering vectors are defined as

\begin{equation*}
\mathbf{e}_{\mathcal{M}}(\overrightarrow{u_{t,k}}, D_k)=\tfrac{1}{\sqrt{N_t}}\begin{pmatrix}
\mathrm{e}^{-\mathrm{j}\frac{2\pi}{\lambda}\Delta_{\mathcal{M},1k}} \\
\vdots \\
\mathrm{e}^{-\mathrm{j}\frac{2\pi}{\lambda}\Delta_{\mathcal{M},N_tk}} 
\end{pmatrix},
\end{equation*}
with
\begin{itemize}[after=]
\setlength\itemsep{0.3em}
\item $\Delta_{\text{PWM},jk} = - \overrightarrow{a_{t,j}}.\overrightarrow{u_{t,k}},$
\item $\Delta_{\text{ParWM},jk} = - \overrightarrow{a_{t,j}}.\overrightarrow{u_{t,k}} + \frac{1}{2D_k}\left[\left\Vert  \overrightarrow{a_{t,j}}\right\Vert^2-( \overrightarrow{a_{t,j}}.\overrightarrow{u_{t,k}})^2\right],$
\item $\Delta_{\text{SWM},jk} = \sqrt{ D_k^2 +2D_k( \overrightarrow{a_{t,j}}.\overrightarrow{u_{t,k}})  + \left \Vert \overrightarrow{a_{t,j}} \right \Vert_2^2  }-D_k.$
\end{itemize}
%\begin{equation*}
%\resizebox{.99\hsize}{!}{$
%\mathbf{e}_{{\text{ParWM}}}(\overrightarrow{u_{t,k}}, D_k)=\tfrac{1}{\sqrt{\llm{N_t}}}\begin{pmatrix}
%\mathrm{e}^{\mathrm{j}\frac{2\pi}{\lambda}\Big(\overrightarrow{a_{t,1}}.\overrightarrow{u_{t,k}}+ \frac{1}{2{D_k}}\big[( \overrightarrow{a_{t,1}}.{\overrightarrow{u_{t,k}}})^2 - \left\Vert  \overrightarrow{a_{t,1}}\right\Vert^2\big]\Big)} \\
%\vdots \\
%\mathrm{e}^{\mathrm{j}\frac{2\pi}{\lambda}\Big(\overrightarrow{a_{t,\llm{N_t}}}.\overrightarrow{u_{t,k}}+ \frac{1}{2{D_k}}\big[( \overrightarrow{a_{t,\llm{N_t}}}.{\overrightarrow{u_{t,k}}})^2 - \left\Vert  \overrightarrow{a_{t,\llm{N_t}}}\right\Vert^2\big]\Big)}
%\end{pmatrix},$}
%\end{equation*}
%and
%\begin{equation*}
%\resizebox{.99\hsize}{!}{$
%\mathbf{e}_{{\text{SWM}}}(\overrightarrow{u_{t,k}}, D_k)=\tfrac{1}{\sqrt{\llm{N_t}}}\begin{pmatrix}
%\mathrm{e}^{\mathrm{j}\frac{2\pi}{\lambda}\big(D_k-\sqrt{ D_k^2 - {2D_k({ \overrightarrow{a_{t,1}}.\overrightarrow{u_{t,k}}})}  + {\left \Vert  \overrightarrow{a_{t,1}} \right \Vert^2}  }\big)} \\
%\vdots \\
%\mathrm{e}^{\mathrm{j}\frac{2\pi}{\lambda}\big(D_k-\sqrt{ D_k^2 - {2D_k({ \overrightarrow{a_{t,\llm{N_t}}}.\overrightarrow{u_{t,k}}})}  + {\left \Vert  \overrightarrow{a_{t,\llm{N_t}}} \right \Vert^2}  }\big)}
%\end{pmatrix}$}.
%\end{equation*}
\llm{These expression are obtained simply by considering $N_r=1$ and $\overrightarrow{a_{r,1}} = \overrightarrow{0}$ in the expressions of $\Delta_{\text{PWM},ij}$, $\Delta_{\text{ParWM},ij}$ and $\Delta_{\text{SWM},ij}$.}
\llm{Note that} the PWM characteristic vectors depend only upon $\overrightarrow{u_{t}}$, they are simply steering vectors. On the other hand, the distance $D$ has an influence on the ParWM and SWM characteristic vectors, since it determines the curvature of the wavefronts.
\section{Validity domains}
\label{sec:validity}
In this section, the goal is to characterize the distance ranges where the different models are describing correctly the channel in the simple LoS case. \llm{The channel is assumed to follow the SWM and the aim is to assess the PWM and ParWM accuracies.} %with any $\llm{N_t}$: since the SWM is always valid here the aim is to quantify the PWM/ParWM accuracy taking the SWM as a reference.\\
\\
\noindent \textbf{Approaches.}
In the literature \cite{Selvan2017, Liu2015} a phase shift difference of \llm{at most} $\frac{\pi}{8}$ \llm{with respect to the SWM phase shift $\frac{2\pi}{\lambda}\Delta_{\text{SWM},jk}$} is used to define the PWM validity. %\llm{Applying it using} \eqref{eq:taylor_expansion} \llm{and keeping only the terms corresponding to} the PWM and the ParWM with $\llm{N_r}=1$, we obtain the boundaries %defining the three validity regions shown in Table \ref{tab:validity_domains}.
\llm{Bounding $\frac{2\pi}{\lambda}|\Delta_{\text{SWM},jk} - \Delta_{\text{PWM},jk}|$ and $\frac{2\pi}{\lambda}|\Delta_{\text{SWM},jk} - \Delta_{\text{ParWM},jk}|$ using the fact that $\left\vert \overrightarrow{a_{t,j}}.\overrightarrow{u_t}\right\vert\leq R_t$, this yields} 
\llm{\begin{itemize}
\setlength\itemsep{0.01em}
\item $D \ge \tfrac{8R_t^2}{\lambda}$ for the PWM, this boundary is often called the Fraunhofer distance \cite{Selvan2017}.
\item $D \ge \sqrt{\tfrac{8R_t^3}{\lambda}}$ for the ParWM, this boundary is sometimes called the Fresnel distance \cite{Selvan2017}.
\end{itemize}}
%\begin{table}[ht]
%  \centering
%  \begin{tabular}{|P{2.8cm}|P{2.8cm}|P{1.2cm}|}
%    \hline
%    PWM & ParWM & SWM \\ \hline
%    \rule{0pt}{14pt}
%     $D \ge \frac{8R_t^2}{\lambda}$ & $D \ge \sqrt{\frac{8R_t^3}{\lambda}}$ & $D \ge 0$ \\ \hline
%  \end{tabular}
%  \caption{Models validity domains (phase-shift approach) }\label{tab:validity_domains}
%\end{table}
\noindent This method has several drawbacks: it considers an arbitrary phase-shift difference of $\frac{\pi}{8}$, does not apply to the overall channel matrix (it is based on individual channel coefficients) and is independent from the relative position of the receiver from the emitter. %: it is conservative. For instance the extreme case of two aligned ULAs implies the PWM to be valid for any distance, instead the phase-shift method could lead to a non-validity conclusion. 
To overcome these drawbacks another metric is introduced \llm{called %thereafter it is referred to as the 
relative model approximation error (rMAE):% and is mathematically defined as
\begin{equation*}
    %\Large
    \label{eq:rmae}
    %\resizebox{.85\hsize}{!}{$
    \text{rMAE}=\frac{\big\Vert\mathbf{h_{}}-\text{proj}_{\mathcal{M}}(\mathbf{h})\big\Vert_{2}^2}{\left\Vert \mathbf{h} \right\Vert_{2}^2},
%    \left \{
%    \begin{array}{rl}
%    \text{proj}_{\mathcal{M}}(\mathbf{U}) \triangleq \underset{\mathbf{X}\in \mathcal{M}_i}{\text{argmin}}\, \left\Vert \mathbf{U} - \mathbf{X} \right\Vert_{2},\\
%    \mathcal{M}_i=\text{PWM, ParWM, SWM},
%    \end{array}
%    \right.$}
\end{equation*}
\normalsize
where $\text{proj}_{\mathcal{M}}(\mathbf{u}) \triangleq \text{argmin}_{\mathbf{x}\in \mathcal{M}}\, \left\Vert \mathbf{u} - \mathbf{x} \right\Vert_{2}$ and $\mathbf{h}$ refers to the true channel (the SWM being taken as the reference, $\mathbf{h}=\mathbf{h}_{\text{SWM}}$). %The projection of $\mathbf{h}$ on a model gives the element minimizing the rMAE: it is the best performance one can achieve with this model.
The rMAE assesses the best approximation of the channel that can be obtained with the considered model.}\\
\noindent \textbf{Setting.}
The objective is to study this new metric \llm{in a single path LoS scenario} varying the array shape (ULA, square UPA), the number of antennas (64, 256), the emitter-receiver distance (from $\lambda$ to $10^{5}\lambda$) and the considered model. A $\frac{\lambda}{2}$ antenna spacing and a single-antenna receiver \llm{located in front of the transmit array (yielding a DoD orthogonal to the array)} are considered. It is to be stressed that the obtained curves are parameterized by the wavelength and thus valid irrespective of the band, even though massive MIMO antenna arrays are more likely to be used at small wavelength (e.g.\ millimeter waves). %The direction of departure is set orthogonal to the array: it is the worst scenario for the PWM.

\begin{figure}[!ht]
    \centering
    \includegraphics[width=\columnwidth]{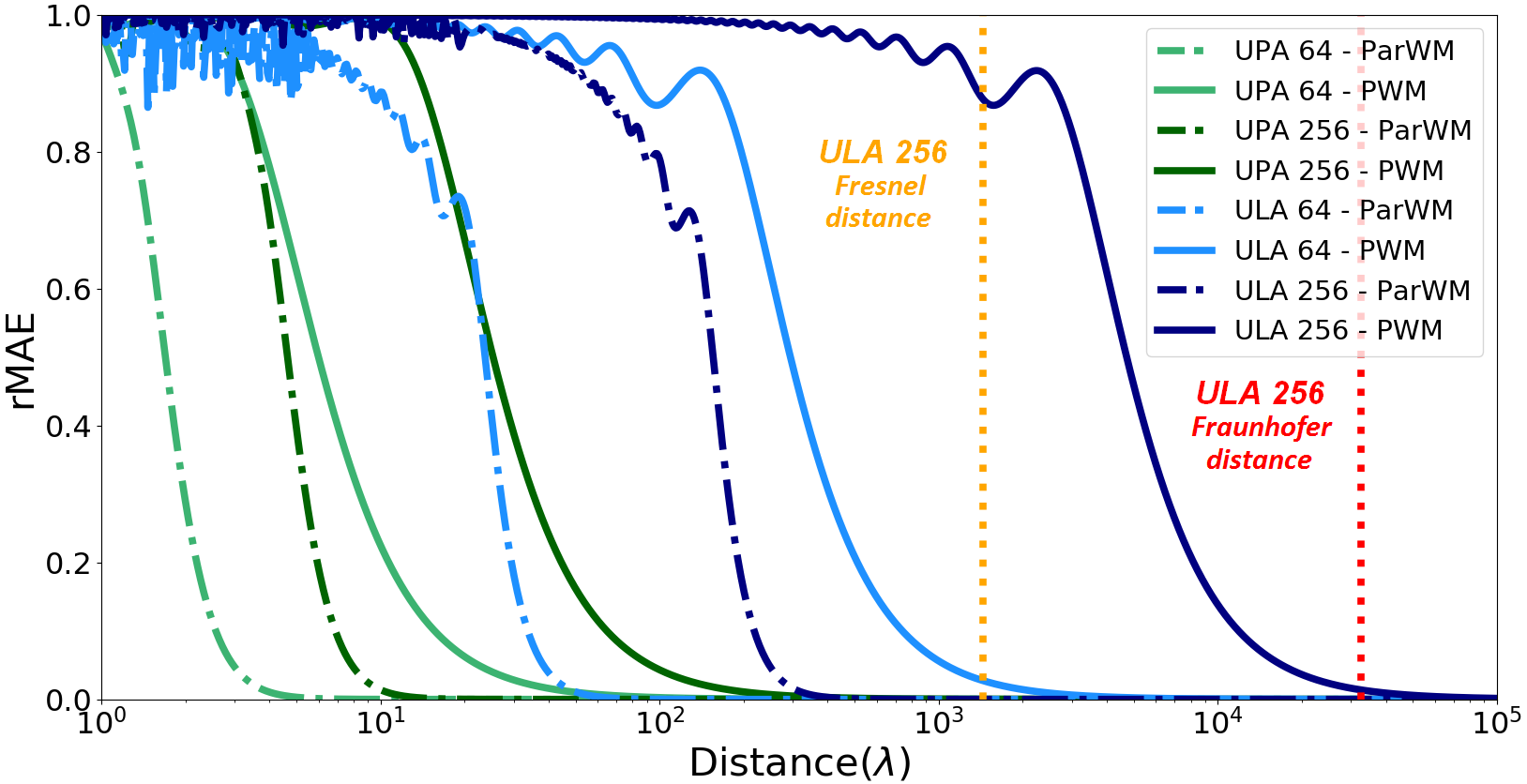}
    \caption{rMAE for the PWM and ParWM varying $D$, $\llm{N_t}$ and the array shape.}
    \label{fig:rMAE} 
\end{figure}

\noindent \textbf{Results.} The figure \ref{fig:rMAE} provides rMAE plots as a function of the normalized distance, expressed on a logarithmic scale, for different configurations. Several comments are in order:
\begin{itemize}[leftmargin=*]
\setlength\itemsep{0mm}
 %Blue (green) curves refer to ULA (UPA). Solid (dotted) lines refer to the PWM (ParWM). Dark (light) curves refer to 256 (64) antennas. 
%\noindent \textbf{Analysis and conclusions.}
\item As expected, at very high distances the rMAE converges to 0 meaning all the models are equivalent \llm{and describe the channel correctly. }%: in particular the PWM describes the channel correctly. 
Nevertheless, % focusing on a ULA with 256 antennas, the error convergence 
\llm{the convergence occurs at distances much smaller for the ParWM than for the PWM: for instance, with a ULA of $256$ antennas}, obtaining $\text{rMAE}<5\%$ at 30GHz requires $D>$ 2.5m for the ParWM and $D>$ 170m for the PWM. In such a setting, the PWM is not suitable, whereas the ParWM is sufficiently accurate. 
% The cell radius, using mmWaves and advanced wireless communication techniques, is expected to be around 200m hence the PWM is no longer suitable to describe mMIMO channels: however an essential conclusion is that the ParWM is sufficiently accurate. 
Obviously considering a ULA with less antennas (here 64) \llm{reduces the critical distances. }%is less critical. 
\item \llm{Another} important observation is that even with many antennas, UPAs do not \llm{incur large} %involve major 
errors: a rMAE of $5\%$ is reached at 1m with 256 antennas. This is simply because for a given number of antennas, \llm{UPAs are much smaller than ULAs.} %shapes induce much smaller arrays ($R_t$ and $R_r$) than linear ones.
 Actually, $R_t$ is proportional to $N_t$ for an ULA and to $\sqrt{N_t}$ for an UPA. %These statements are in line with the table \ref{tab:validity_domains} where the limit distances are increasing with the array size.
\item Finally, the yellow (red) vertical line gives the distance boundary for the ParWM (PWM) computed using phase shifts %in table \ref{tab:validity_domains} 
with a ULA of 256 antennas. \llm{Beyond} %above 
\llm{this} %the yellow (red) vertical 
line which corresponds to $D=$15m ($D=$320m) %defining the conservative validity domains of the ParWM (PWM) 
at 30GHz, %for the ULA with 256 antennas, 
the error is \llm{as expected} negligible.%: it is coherent.
% The simulated scenario ($\overrightarrow{u_t}$ orthogonal to the array) is actually the worst for the PWM and the most favorable for the ParWM which explains why the ParWM error converges before 15m. Other scenarios induce the ParWM error to converge closer to it. Still the parabolic wave model appears as a very good approximation of the spherical one.\\
\end{itemize}

\indent This study highlights the limits of the PWM at short distances \llm{in a novel way, considering the channel matrix globally.} On the other hand, the ParWM is shown to be accurate at such short distances, under which users are likely to be present in practical situations. %, which is} different%ly 
%from what has been proposed so far in the literature.
 Additionally, it clearly shows that ULAs are more challenging for the PWM than UPAs \llm{for which the PWM is accurate from short distances,} which is in line with theory.%The coherency of the approach with respect to the traditional phase-shift one is underlined.

\section{Estimation algorithms}
\label{sec:typestyle}
%Acquiring the channel state information (CSI) is essential to optimize the capacity of mMIMO systems. To achieve it a training-based strategy is commonly adopted: it consists in transmitting $n_t$ pilot symbols at each of the $n_s$ estimation time slots. It yields the transmission equation which can be vectorized and particularized to a single-antenna receiver as

In the previous section, the intrinsic accuracy of models was assessed. Let us now study how to estimate the channel using these models, based on noisy observations. Indeed, channel state information (CSI) is essential to optimize the capacity of mMIMO systems. Consider a training based estimation strategy in which $N_s$ noisy linear measurements of the channel are taken:
\begin{equation*}
    \label{eq:mimo_estimation}
%   \mathbf{Y}=\mathbf{H}\mathbf{X}+\mathbf{N} \overset{\text{vec}(\cdot)}{\underset{n_r=1}{\Rightarrow}} \mathbf{y}=\mathbf{X}^T\mathbf{h}+\mathbf{n} \underset{\mathbf{X}=\mathbf{Id}}{\Rightarrow} = \mathbf{y}=\mathbf{h}+\mathbf{n}
\mathbf{y}=\mathbf{X}\mathbf{h}+\mathbf{n},
\end{equation*}
\noindent where $\mathbf{y} \in \mathbb{C}^{N_s}$ is the observation, $\mathbf{X} \in \mathbb{C}^{N_s \times N_t}$ is the obervation matrix (pilot symbols) and $\mathbf{n}\in \mathbb{C}^{N_s}$ is the noise vector. %contains the received pilot symbols altered by the channel and the noise. The second step is obtained assuming $\mathbf{X}=\mathbf{Id}$: since this paper is focusing on the models and not on the training matrix $\mathbf{X}$, this choice is convenient and reasonable. Note that the hybrid architecture notations are not used here, they are available in \cite{Lemagoarou2018}. 
Under the additive white Gaussian noise (AWGN) assumption ($\mathbf{n}\sim\mathcal{CN}(0,\sigma^2\mathbf{Id})$), a classical estimation technique is the maximum likelihood (ML), which according to the considered models \eqref{eq:general_channel_matrix} can be written as %the optimization problem
%\begin{equation}
%    \label{eq:ML}
%\underset{\mathbf{h}_\mathcal{M}}{\text{argmin}}\;\big\Vert \mathbf{y} -  \mathbf{X}\mathbf{h}_\mathcal{M}\big\Vert_2^2,
%\end{equation}
%where $\mathbf{h}_\mathcal{M}$ depends on the parameters according to the considered physical model as in \eqref{eq:general_channel_matrix}.
% Using a physical model the ML consists in finding the parameters maximizing the observations probability. For a given path it is possible to perform likelihood concentration
%on $\rho$ and $\phi$ hence spare two parameters to estimate.
\begin{equation*}
    \label{eq:ML2}
\underset{\mathbf{E},\boldsymbol{\alpha}}{\text{minimize}}\;\big\Vert \mathbf{y} -  \mathbf{X}\mathbf{E}\boldsymbol{\alpha}\big\Vert_2^2,\quad \hat{\mathbf{h}} \leftarrow \mathbf{E}\boldsymbol{\alpha}
\end{equation*}
where $\mathbf{E} \triangleq \left(\mathbf{e}_{{\mathcal{M}}}(\overrightarrow{u_{t,1}}, D_1),\dots,\mathbf{e}_{{\mathcal{M}}}(\overrightarrow{u_{t,p}}, D_p)\right)$, $\boldsymbol{\alpha}\triangleq \sqrt{N_t}(\rho_1\mathrm{e}^{\mathrm{j}\phi_1} ,\allowbreak \dots , \rho_p\mathrm{e}^{\mathrm{j}\phi_p})^T$ and $\hat{\mathbf{h}}$ is the channel estimate. Note that given $\mathbf{E}$, the optimal vector $\boldsymbol{\alpha}$ can be obtained as the solution of a least squares problem as $\boldsymbol{\alpha}_{\text{opt}} = (\mathbf{E}^H\mathbf{X}^H\mathbf{XE})^{-1}\mathbf{E}^H\mathbf{X}^H\mathbf{y}$, so that in the end channel estimation amounts to find an optimal $\mathbf{E}$, i.e.\ an optimal set of $p$ characteristic vectors $\left\{\mathbf{e}_{{\mathcal{M}}}(\overrightarrow{u_{t,1}}, D_1),\dots,\mathbf{e}_{{\mathcal{M}}}(\overrightarrow{u_{t,p}}, D_p)\right\}$.

\noindent \textbf{Greedy strategy for the PWM.} Looking for the $p$ vectors jointly yields a very complex optimization problem. Instead, greedy strategies have been proposed in the PWM case which consist in building a dictionary of characteristic (steering) vectors corresponding to $N_{\overrightarrow{u_t}}$ DoDs and applying a sparse recovery algorithm such as orthogonal matching pursuit (OMP) \cite{Mallat1993,Tropp2007,Bajwa2010}. This amounts to estimate the paths one by one, i.e.\ building the matrix $\mathbf{E}$ column by column. Denoting $\mathbf{E}^{(k)} \triangleq \left(\mathbf{e}_{{\text{PWM}}}(\overrightarrow{u_{t,1}}, D_1),\dots,\mathbf{e}_{{\text{PWM}}}(\overrightarrow{u_{t,k}}, D_k)\right)$ the state of the matrix $\mathbf{E}$ at the $k$-th iteration, the optimal vector $\boldsymbol{\alpha}^{(k)} \gets (\mathbf{E}^{(k)H}\mathbf{X}^H\mathbf{X}\mathbf{E}^{(k)})^{-1}\mathbf{E}^{(k)H}\mathbf{X}^H\mathbf{y}$ is computed so that a residual $\mathbf{r}^{(k+1)} \gets \mathbf{y} - \mathbf{XE}^{(k)} \boldsymbol{\alpha}^{(k)}$ is used at the next iteration. The actual choice of the $k$-th column of $\mathbf{E}$ is done by finding

% To estimate multipath channel a solution is to combine ML and an OMP algorithm: at each iteration an additional path is estimated, all the paths coefficients are updated and their contributions are removed from the observations so that we can proceed with the next path. To perform the estimation naive strategies consist in testing $N_{\overrightarrow{u_t}}$ steering vectors for the PWM, $N_{\overrightarrow{u_t}}N_D$ steering vectors for the ParWM or the SWM and pick the one that fits best the observations. For the path $k$ the optimization problem is 

\begin{equation}\tag{$S_{\text{PWM}}$}
\overrightarrow{u_{t,k}}  \leftarrow  \underset{\overrightarrow{u_t}}{\text{argmax}}\frac{\big\vert \mathbf{r}^{(k)H}\mathbf{Xe}_{\text{PWM}}(\overrightarrow{u_t}) \big\vert}{\big\Vert \mathbf{Xe}_{\text{PWM}}(\overrightarrow{u_t}) \big\Vert_2},
\label{eq:PWM_strategy}
\end{equation}
among the $N_{\overrightarrow{u_t}}$ test directions. The complexity of this strategy is dominated by the computation of $N_{\overrightarrow{u_t}}$ inner products in $\mathbb{C}^{N_t}$.

\noindent \textbf{Adaptation to SWM and ParWM.} One possible, although naive way to handle the SWM and ParWM is to adopt the same strategy except that the choice of the $k$-th column of $\mathbf{E}$ is done by finding 
\begin{equation}\tag{$S_{\text{joint}}$}
\overrightarrow{u_{t,k}},D_k  \leftarrow \underset{\overrightarrow{u_t},D}{\text{argmax}}\frac{\big\vert \mathbf{r}^{(k)H}\mathbf{Xe}_{\mathcal{M}}(\overrightarrow{u_t},D) \big\vert}{\big\Vert \mathbf{Xe}_{\mathcal{M}}(\overrightarrow{u_t},D) \big\Vert_2},
\label{eq:joint}
\end{equation}
%This is the approach taken by classical greedy approaches based on MP/OMP where a dictionary is built.
%\noindent where $\mathbf{r}^{(i)H}$ corresponds to the observations from which the first $k-1$ paths have been subtracted. 
%\balance
\flushcolsend 
where $\mathcal{M}$ stands for ParWM or SWM. Testing jointly $N_{\overrightarrow{u_t}}$ directions and $N_D$ distances, solving this optimization problem amounts to test $N_{\overrightarrow{u_t}}N_D$ vectors $\mathbf{e}_{\mathcal{M}}(\overrightarrow{u_t},D)$.  This yields a complexity dominated by the computation of $N_{\overrightarrow{u_t}}N_D$ inner products in $\mathbb{C}^{N_t}$.
%Note that the distance $D$ is not estimated for the PWM since it is not a parameter of the model whereas for the ParWM and SWM the DoD and the distance are estimated jointly. As described later this naive strategy implies a very high complexity to account for the wave sphericity. Instead what we propose is to estimate sequentially the DoD and the distance. The DoD is estimated using the PWM through

Another possibility is to estimate the direction and distance sequentially, assuming an infinite distance during the direction determination (which amounts to consider the PWM), yielding

%\noindent and the distance using either the ParWM or the SWM through

\begin{equation}\tag{$S_{\text{seq}}$}
%\begin{array}{l}
%\begin{dcases}
\begin{aligned}
&\overrightarrow{u_{t,k}}  \leftarrow  \underset{\overrightarrow{u_t}}{\text{argmax}}\frac{\big\vert \mathbf{r}^{(k)H}\mathbf{Xe}_{\text{PWM}}(\overrightarrow{u_t}) \big\vert}{\big\Vert \mathbf{Xe}_{\text{PWM}}(\overrightarrow{u_t}) \big\Vert_2}, \\
&D_k  \leftarrow \underset{D}{\text{argmax}}\frac{\big\vert \mathbf{r}^{(k)H}\mathbf{Xe}_{\mathcal{M}}(\overrightarrow{u_{t,k}},D) \big\vert}{\big\Vert \mathbf{Xe}_{\mathcal{M}}(\overrightarrow{u_t},D) \big\Vert_2}.
\end{aligned}
%\end{dcases}
%\end{array}
\label{eq:sequential}
\end{equation}
Testing $N_{\overrightarrow{u_t}}$ directions and $N_D$ distances, solving this optimization problem amounts to test $N_{\overrightarrow{u_t}}$ vectors $\mathbf{e}_{PWM}(\overrightarrow{u_{t}},D)$ and then $N_D$ vectors $\mathbf{e}_{\mathcal{M}}(\overrightarrow{u_{t,k}},D)$.  This yields a complexity dominated by the computation of only $N_{\overrightarrow{u_t}} + N_D$ inner products in $\mathbb{C}^{N_t}$.

\noindent \textbf{Preliminary experiment.} Let us compare empirically the three aforementioned strategies. To do so, consider a base station equipped with an ULA with 256 antennas and a single-antenna user terminal located $D=20$m away from the base station at a random angle $\beta \in [-60^\circ,60^\circ]$ with respect to the direction orthogonal to the array. %From the simulations performed previously the ULA appeared as the most critical shape hence it is the one used hereafter with 256 antennas. 
The channel is randomly generated and composed of a LoS path and a random number of NLoS paths drawn uniformly at random between 0 and 5, the number of path is unknown for estimation. Reflectors positions are randomly generated so that the length of reflected paths is no longer than $2D$: they are modeled by perfect planes inducing uniformly distributed phase shifts between $0$ and $2\pi$ and Rayleigh distributed attenuations with $\sigma=0.3$. The SNR is set to $10$dB and $\mathbf{X}=\mathbf{Id}$ is taken (the objective here is to assess the various estimation strategies, not a specific pilot configuration). $N_{\overrightarrow{u_t}}=300$ test directions uniformly sampling $[0,\pi]$, and $N_D=20$ test distances logarithmically distributed between $1$m and $1$km are tested.

\begin{figure}[!ht]
    \centering
    \includegraphics[width=\columnwidth]{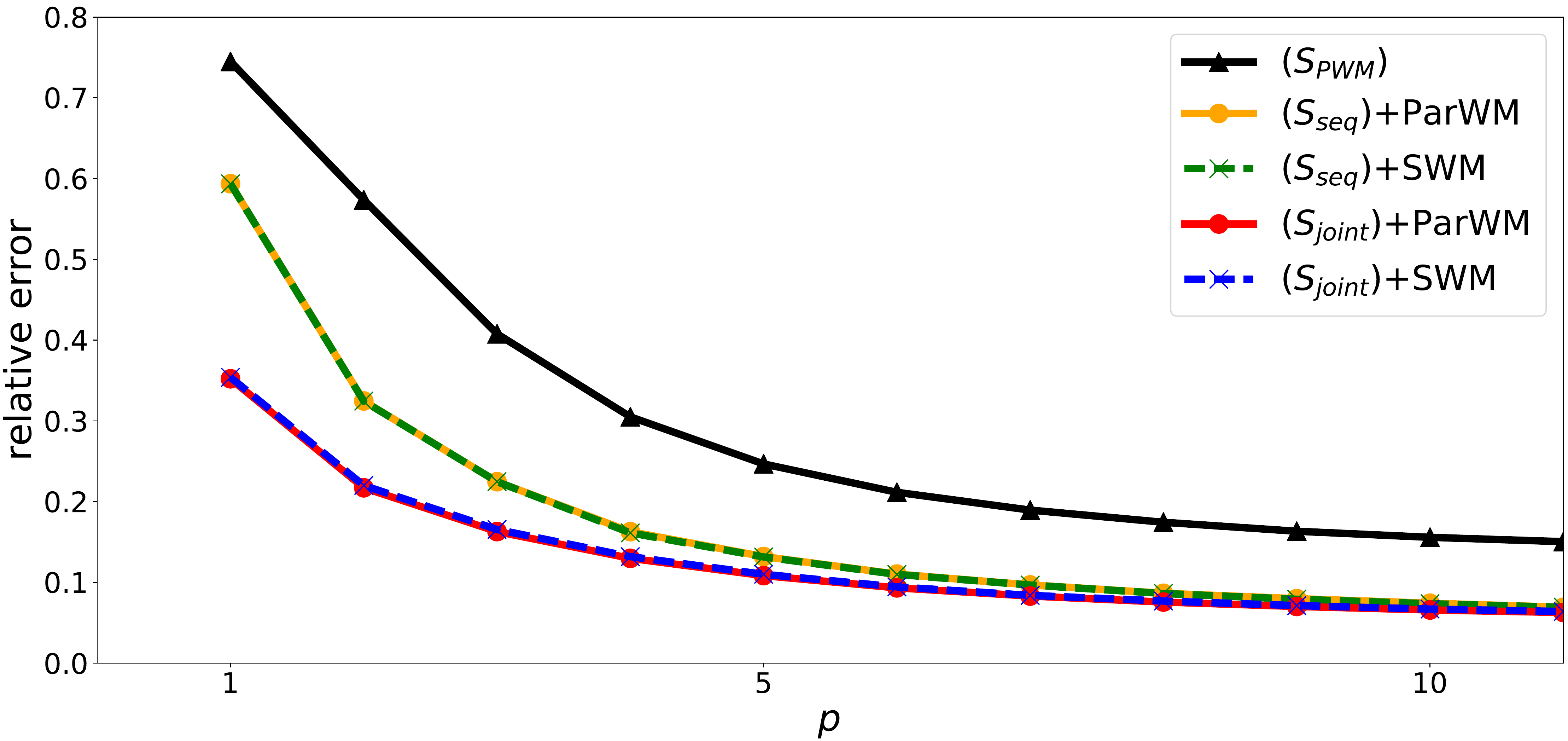}

    %~\vspace{-5mm}
    
    \caption{Comparison of the estimation algorithms.}
    \label{fig:estimation} 
\end{figure}

%~\vspace{-5mm}

\noindent \textbf{Results.} The experiment results for a varying number of estimated paths $p$ are shown in figure~\ref{fig:estimation}. The metric used to assess the estimation strategies is the relative error $\frac{\Vert\mathbf{h_{}}-\hat{\mathbf{h}}\Vert_{2}^2}{\left\Vert {\mathbf{h}} \right\Vert_{2}^2}$, and averages over 100 realizations are shown. For all methods, taking $p$ around five allows to get reasonable estimates (around 10\% relative error for methods taking into account curvature and around 25\% for \eqref{eq:PWM_strategy}). Taking $p$ too small leads to oversimplified channel estimates while taking $p$ too large requires more computations and does not improve the estimates, due to the noise level. This is a bias-variance trade-off \cite{Lemagoarou2018b}. Let us now compare the different methods:
\begin{itemize}[leftmargin=*]
\setlength\itemsep{0.0em}
\item First of all, as expected, the strategies \eqref{eq:joint} and \eqref{eq:sequential} are much better than \eqref{eq:PWM_strategy}. this is because they take into account the wavefronts curvature.
\item Moreover, \eqref{eq:sequential} is almost as good as \eqref{eq:joint} (as soon as $p>4$), which shows its interesting potential. It is indeed much more computationally efficient than \eqref{eq:joint} (at least fifteen times faster in the tested configuration on a laptop with an Intel(R) Core(TM) i7-3740QM CPU @ 2.70 GHz). %\eqref{eq:sequential} is a very good trade-off: it is almost as computationally efficient as \eqref{eq:PWM_strategy} and almost as accurate as \eqref{eq:joint}. 
\item Finally, ParWM and SWM are equivalent in the considered setting, despite ParWM being simpler (it does not involve square roots). This is interesting for a hardware implementation in which complex operations are preferably avoided.
\end{itemize}

%The relative error of the alternative algorithm converges faster than the PWM: for most of the iterations its relative error is at least twice lower. However it converges slower than the naive algorithm based on ParWM and SWM: it is the cost of the complexity reduction. Our proposal appears as a good performance-complexity trade-off. Additionally the ParWM and the SWM are, using either the naive or the alternative proposal, equivalent in terms of performance, which proves even more the interest of the ParWM.

%~\vspace{-5mm}

\section{Conclusions and Perspectives}
In this paper three physical channel models applicable to massive MIMO and any type of antenna array have been studied in an unified way: the well-known PWM, the SWM and the novel ParWM which yields an interesting accuracy-complexity trade-off. The models accuracies have been assessed, underlining the PWM limitations in particular when large ULAs are considered, which is plausible in practical massive MIMO scenarios. % The parabolic wave model presents an interest in situations for which the plane wave model is not sufficient (i.e. ULA with many antennas) to describe the channel: such cases are plausible in practical massive MIMO scenarios.
Two estimation algorithms taking the wavefronts curvature (SWM or ParWM) into account have been proposed and compared to the classical PWM approach. In particular, a computationally efficient strategy in which the DoD and distance are estimated sequentially has been proposed, showing promising results.

In the future, assessing precisely the influence of the distance in order to determine where the ParWM/SWM is necessary would be very useful. Obviously, a more extensive experimental evaluation of the proposed algorithms should also be undertaken. Moreover, generalizing the proposed channel estimation methods to multi-antenna receivers in NLoS scenarios would be of interest.

\newpage
\newpage
\bibliographystyle{IEEEtran}
%\bibliographystyle{IEEEbib}
%\bibliography{strings,refs}
\bibliography{biblio}

\begin{thebibliography}{10}
\providecommand{\url}[1]{#1}
\csname url@samestyle\endcsname
\providecommand{\newblock}{\relax}
\providecommand{\bibinfo}[2]{#2}
\providecommand{\BIBentrySTDinterwordspacing}{\spaceskip=0pt\relax}
\providecommand{\BIBentryALTinterwordstretchfactor}{4}
\providecommand{\BIBentryALTinterwordspacing}{\spaceskip=\fontdimen2\font plus
\BIBentryALTinterwordstretchfactor\fontdimen3\font minus
  \fontdimen4\font\relax}
\providecommand{\BIBforeignlanguage}[2]{{%
\expandafter\ifx\csname l@#1\endcsname\relax
\typeout{** WARNING: IEEEtran.bst: No hyphenation pattern has been}%
\typeout{** loaded for the language `#1'. Using the pattern for}%
\typeout{** the default language instead.}%
\else
\language=\csname l@#1\endcsname
\fi
#2}}
\providecommand{\BIBdecl}{\relax}
\BIBdecl

\bibitem{Bjornson2017}
E.~Bj{\"o}rnson, J.~Hoydis, L.~Sanguinetti \emph{et~al.}, ``Massive {MIMO}
  networks: Spectral, energy, and hardware efficiency,'' \emph{Foundations and
  Trends in Signal Processing}, vol.~11, no. 3-4, pp. 154--655, 2017.

\bibitem{Bjornson2016}
E.~Bj{\"o}rnson, E.~G. Larsson, and T.~L. Marzetta, ``Massive {MIMO}: Ten myths
  and one critical question,'' \emph{IEEE Communications Magazine}, vol.~54,
  no.~2, pp. 114--123, 2016.

\bibitem{Larsson2014}
E.~G. Larsson, O.~Edfors, F.~Tufvesson, and T.~L. Marzetta, ``Massive {MIMO}
  for next generation wireless systems,'' \emph{IEEE Communications Magazine},
  vol.~52, no.~2, pp. 186--195, 2014.

\bibitem{Lu2014}
L.~Lu, G.~Y. Li, A.~L. Swindlehurst, A.~Ashikhmin, and R.~Zhang, ``An overview
  of massive {MIMO}: Benefits and challenges,'' \emph{IEEE Journal of Selected
  Topics in Signal Processing}, vol.~8, no.~5, pp. 742--758, 2014.

\bibitem{Hampton2013}
J.~R. Hampton, \emph{Introduction to {MIMO} Communications}.\hskip 1em plus
  0.5em minus 0.4em\relax Cambridge {University Press}, 2013.

\bibitem{Rusek2013}
F.~Rusek, D.~Persson, B.~K. Lau, E.~G. Larsson, T.~L. Marzetta, O.~Edfors, and
  F.~Tufvesson, ``Scaling up {MIMO}: Opportunities and challenges with very
  large arrays,'' \emph{IEEE Signal Processing Magazine}, vol.~30, no.~1, pp.
  40--60, 2013.

\bibitem{Swindlehurst2014}
A.~L. Swindlehurst, E.~Ayanoglu, P.~Heydari, and F.~Capolino, ``Millimeter-wave
  massive {MIMO}: the next wireless revolution?'' \emph{IEEE Communications
  Magazine}, vol.~52, no.~9, pp. 56--62, 2014.

\bibitem{Rappaport2013}
T.~S. Rappaport, S.~Sun, R.~Mayzus, H.~Zhao, Y.~Azar, K.~Wang, G.~N. Wong,
  J.~K. Schulz, M.~Samimi, and F.~Gutierrez, ``Millimeter wave mobile
  communications for {5G} cellular: It will work!'' \emph{IEEE Access}, vol.~1,
  pp. 335--349, 2013.

\bibitem{Telatar1999}
E.~Telatar, ``Capacity of multi-antenna {Gaussian} channels,'' \emph{European
  Trans. on Telecommunications}, vol.~10, no.~6, pp. 585--595, 1999.

\bibitem{Lemagoarou2018}
L.~{Le Magoarou} and S.~Paquelet, ``Parametric channel estimation for massive
  {MIMO},'' in \emph{IEEE Statistical Signal Processing Workshop (SSP)}, 2018.

\bibitem{Selvan2017}
K.~T. Selvan and R.~Janaswamy, ``Fraunhofer and fresnel distances: Unified
  derivation for aperture antennas.'' \emph{IEEE Antennas and Propagation
  Magazine}, vol.~59, no.~4, pp. 12--15, 2017.

\bibitem{Cheng2017}
X.~Cheng and Y.~He, ``Geometrical model for massive {MIMO} systems,'' in
  \emph{2017 IEEE 85th Vehicular Technology Conference (VTC Spring)}, 2017, pp.
  1--6.

\bibitem{Tamaddondar2017}
M.~M. Tamaddondar and N.~Noori, ``Plane wave against spherical wave assumption
  for non-uniform linear massive {MIMO} array structures in {LoS} condition,''
  in \emph{2017 Iranian Conference on Electrical Engineering (ICEE)}, 2017, pp.
  1802--1805.

\bibitem{Liu2016}
L.~Liu, D.~W. Matolak, C.~Tao, Y.~Li, B.~Ai, and H.~Chen, ``Channel capacity
  investigation of a linear massive {MIMO} system using spherical wave model in
  {LoS} scenarios,'' \emph{Science China Information Sciences}, vol.~59, no.~2,
  pp. 1--15, 2016.

\bibitem{Liu2015}
L.~Liu, D.~W. Matolak, C.~Tao, Y.~Lu, and H.~Chen, ``Far region boundary
  definition of linear massive {MIMO} antenna arrays,'' in \emph{2015 IEEE 82nd
  Vehicular Technology Conference (VTC2015-Fall)}, 2015, pp. 1--6.

\bibitem{Zhou2015}
Z.~Zhou, X.~Gao, J.~Fang, and Z.~Chen, ``Spherical wave channel and analysis
  for large linear array in {LoS} conditions,'' in \emph{Globecom Workshops,
  2015 IEEE}.\hskip 1em plus 0.5em minus 0.4em\relax IEEE, 2015, pp. 1--6.

\bibitem{Bohagen2009}
F.~Bohagen, P.~Orten, and G.~E. Oien, ``On spherical vs. plane wave modeling of
  line-of-sight {MIMO} channels,'' \emph{IEEE Transactions on Communications},
  vol.~57, no.~3, pp. 841--849, 2009.

\bibitem{Jiang2005}
J.-S. Jiang and M.~A. Ingram, ``Spherical-wave model for short-range {MIMO},''
  \emph{IEEE Transactions on Communications}, vol.~53, no.~9, pp. 1534--1541,
  2005.

\bibitem{Mallat1993}
S.~Mallat and Z.~Zhang, ``Matching pursuits with time-frequency dictionaries,''
  \emph{Signal Processing, IEEE Transactions on}, vol.~41, no.~12, pp.
  3397--3415, Dec 1993.

\bibitem{Tropp2007}
J.~Tropp and A.~Gilbert, ``Signal recovery from random measurements via
  orthogonal matching pursuit,'' \emph{Information Theory, IEEE Transactions
  on}, vol.~53, no.~12, pp. 4655--4666, Dec 2007.

\bibitem{Bajwa2010}
W.~U. Bajwa, J.~Haupt, A.~M. Sayeed, and R.~Nowak, ``Compressed channel
  sensing: A new approach to estimating sparse multipath channels,''
  \emph{Proceedings of the IEEE}, vol.~98, no.~6, pp. 1058--1076, June 2010.

\bibitem{Lemagoarou2018b}
L.~{Le Magoarou} and S.~Paquelet, ``Bias-variance tradeoff in {MIMO} channel
  estimation,'' \emph{arXiv:1804.07529}, 2018.

\end{thebibliography}
\end{document}